\documentclass[12pt,twoside]{article}   
\usepackage{epsfig}
\title{\bf Does Sgr A* Have an Event Horizon or an Intrinsic Magnetic Moment?}

\author{Stanley L. Robertson\footnote{Physics Dept., Southwestern Oklahoma State University,
Weatherford, OK 73096, USA (stan.robertson@swosu.edu)} and Darryl
J. Leiter\footnote{Visiting Scientist, National Radio Astronomy
Observatory, Charlottesville, VA 22903, USA (dleiter@nrao.edu)}}

\begin{document}
\maketitle

\begin{abstract}In previous work we have presented evidence for the
existence of intrinsic magnetic moments in black hole candidates.
We later developed a general relativistic, magnetospheric
eternally collapsing object (MECO) model for black hole candidates
and showed that the model is consistent with broad band spectral
and luminosity characteristics and accounts for the radio/x-ray
luminosity correlations of both galactic black hole candidates
(GBHC) and active galactic nuclei (AGN). Since magnetic moments
are forbidden attributes for black holes, the MECO model has the
potential to test whether black hole candidates are actual black
holes. We show here that the MECO model has the advantage of being
able to: a) satisfy the luminosity constraints that have been
claimed as proof of an event horizon for Sgr A*,  b) reconcile the
low bolometric luminosity with the expected Bondi accretion rate
for Sgr A* by means of a magnetic propeller driven outflow,  c)
account for the Sgr A* NIR and x-ray luminosities, the general
characteristics of its broad band spectrum, and the sequence of
flares in different spectral ranges as well as the pattern of its
observed orthogonal polarizations. We also include specific
predictions for images that may be obtained in sub-millimeter to
NIR wavelengths in the near future. High resolution images in
radio frequencies should be elongated in an equatorial outflow
plane, while high resolution images in shorter infrared
wavelengths should be elongated along an orthogonal, magnetic
polar inflow axis (generally N-S). Since the emissions in these
shorter wavelengths are confined to a narrow axial inflow cone,
and radio frequencies are generated primarily at greater distances
in the equatorial outflow, there would be no uniform background to
provide a silhouette image of a dark central object. Additional
future tests for the presence of an intrinsic magnetic moment for
Sgr A* will require global solutions for electron density and
magnetic field distributions in a Bondi accretion flow into a
compact, rotating magnetic dipole. These will provide for
definitive tests in the form of detailed calculations of spectral
and spatial luminosity distributions and polarization maps for
direct comparison with high resolution images of Sgr A*.
\end{abstract}

\noindent {\it Key words:} accretion, accretion disks--black hole
physics--Galaxy : center--gravitation--infrared: general--magnetic
fields


\section{Introduction}
The existence of black holes is accepted as a fact of the cosmos
by many astrophysicists. The evidence for the existence of massive
objects that may be compact enough to be black holes is strong,
however there is as yet no direct evidence of any mass that is
contained within its Schwarzschild radius and even that would not
necessarily prove the object to be a black hole. Supermassive
compact objects have been found in the nuclei of most galaxies and
objects of stellar mass are abundant within our own and other
galaxies. While these compact objects are commonly called black
holes no compelling observational evidence of an event horizon,
the quintessential feature of a black hole, has yet been found. It
has been pointed out previously (Abramowicz, Kluzniak \& Lasota
2002) that it would be very difficult to prove the existence of
event horizons if objects with large surface gravitational
redshift exist. Extreme redshifts could make such objects nearly
as dark as a black hole (see Appendix A). If not different from a
black hole in some other way, it would not seem to matter much
whether or not the black hole candidates are actual black holes.
On the other hand, nearly every object of stellar mass or greater
is known to be magnetic to some degree. Whatever their origins,
magnetic fields and their synchrotron radiations are ubiquitous
among the black hole candidates. Instead of possessing event
horizons, some of these compact objects might be found to be both
intrinsically magnetic and highly redshifted. Since magnetic
moments are forbidden attributes for black holes, there would be
important consequences for astrophysics if some of the black hole
candidates were found to possess them. Already some evidence for
the existence of magnetic moments has been presented for both GBHC
and AGN (Robertson \& Leiter 2002, 2004, 2006 Schild, Leiter \&
Robertson 2006, 2008). Others have reported evidence for very
strong magnetic fields in GBHC, e.g., a field in excess of $10^8$
G has been found at the base of the jets of GRS 1915+105 (Gliozzi,
Bodo \& Ghisellini 1999, Vadawale, Rao \& Chakrabarti 2001).

These findings provided motivation for the development of a fully
general relativistic model of a gravitationally compact,
intrinsically magnetic, eternally collapsing object that we call a
MECO (Leiter \& Robertson 2003, Mitra 2006a,b,c, Robertson \&
Leiter 2003, 2004, 2006, hereafter RL03, RL04, RL06). A MECO
avoids rapid collapse to a black hole state by radiating away its
mass-energy at an Eddington limit rate. It is characterized by
both an extreme redshift and a strong intrinsic magnetic moment
(see Appendix A - D). The large redshift accounts for the low
quiescent surface luminosities of GBHC and AGN and their extremely
long (many Hubble times) radiative lifetimes. The MECO model has
been shown (McClintock, Narayan \& Rybicki 2004) to be consistent
with the lower limit on the quiescent emission from the GBHC XTE
J1118+480. We show here that it is also consistent with the low
luminosity of Sgr A*.

The MECO model for disk accreting GBHC and AGN accounts for the
existence of their low/hard and high/soft x-ray spectral states.
The rotating magnetic moments provide a robust universal magnetic
propeller mechanism for spectral state switches (Ilarianov \&
Sunyaev 1975, Campana et al., 1998, 2002, RL02, RL03). The
high/soft $\rightarrow$ low/hard transition marks the start of a
magnetic propeller regime, the end of accreting plasma being able
to penetrate inside the corotation radius and the beginning of a
low state jet outflow. The MECO model correctly predicts (RL04)
that this transition occurs at about $\sim 0.02$ of Eddington
limit luminosity for both GBHC and AGN. The luminosities at the
transition and in quiescence have permitted the determinations of
the magnetic moments and spin rates for MECO-GBHC (RL02, RL06, and
Eqs 6 \& 10, Table 1). Spectral state switches are common to dwarf
novae, the neutron stars and GBHC of low mass x-ray binary systems
and AGN. They are signatures of intrinsic magnetism, however they
have not yet been accepted as such because it is generally
believed that the black hole candidates are actual black holes and
cannot possess magnetic moments.

MECO interacting with accretion disks, as shown in RL04, can
produce low/hard state jet outflows with correlated radio and
x-ray emissions ($L_R \propto L_x^{2/3}$) in accord with
observations (Gallo, Fender \& Pooley 2003, Markoff et al. 2003,
Falcke, K\"{o}rding \& Markoff 2004, Maccarone, Gallo \& Fender
2003). To account for the similar correlations for AGN and neutron
star or dwarf nova systems, it was necessary to explore the mass
scaling relations for the MECO model (RL04). These MECO mass
scaling relationships, including those associated with the
magnetic moments and spin rates that have been able to account for
the Radio/X-ray luminosity correlations of AGN and GBHC (RL04),
are listed in the right hand column of Table 1. It is a remarkable
fact that no further adjustments to these MECO mass scaling
parameters are needed for the successful application of the MECO
model to the case of Sgr A*. They also account for the newly
discovered quasar accretion structures (Schild, Leiter \&
Robertson 2006, 2008) revealed by microlensing observations of the
quasars Q0957+561 and Q2237+0305. The observed structures are
consistent with the strongly magnetic MECO model but do not accord
with standard thin disk models of accretion flows into a black
hole. Since even the nearest GBHC are much too small to be
resolved in the detail shown by the microlensing techniques which
were used in the study of Q0957 and Q2237, Sgr A* is likely the
only remaining black hole candidate for which resolved images
might reveal whether or not it possesses a magnetic moment. For
this reason it is important that it be tested.

Because much of the spectrum of Sgr A* appears to originate in
synchrotron-cyclotron radiation, a serious test of the hypothesis
that Sgr A* might possess an intrinsic magnetic moment will
require global solutions for the magnetic field and electron
density distributions for some kind of accretion flow. While this
will require future detailed simulations that are beyond the scope
of the present work, we show in this paper that analytic methods
can be used to give here a good accounting of the physical
properties of the radio/NIR and X-ray spectral characteristics,
luminosities and polarizations that have been observed for Sgr A*.
We find that we must consider Bondi accretion for a quiescent MECO
and then show that we can reconcile the observed low luminosity of
Sgr A* with the expected Bondi accretion flow rate.

For the discussion to follow and also for the convenience of
readers who might wish to relate MECO properties to observations
of other objects, we have tabulated a number of useful relations
in Table 1. Many of the parameters are given in terms of quiescent
x-ray luminosity $L_q$, or the luminosity, $L_c$, at the
transition high/soft $\rightarrow$ low/hard state since these are
often measurable quantities. Some new developments, minor
corrections and general features for the MECO model are presented
in Appendixes A - E.

\begin{table*}
\tiny

\begin{center}
\caption{MECO Model Equations}
\end{center}
\begin{tabular}{lll} \hline
MECO Physical Quantity & Equation & (Scaling, $m$ in $M_\odot$)\\
\hline 1.~~~Surface Redshift - (RL06 2) & $1+z_s = 5.67\times 10^7
m^{1/2}$ &
$m^{1/2}~$\\
2.~~~Quiescent Surface Luminosity $L_\infty$ - (RL06 29) &
$L_\infty=1.26\times 10^{38} m/(1+z_s)$ erg/s & $m^{1/2}~$
\\
3.~~~Quiescent Surface Temp $T_\infty$ - (RL06 31) &
$T_\infty=2.3\times 10^7/[m(1+z_s)]^{1/4}=2.65\times 10^{5}
m^{-3/8}$ K & $m^{-
3/8}$\\

4.~~~Photosphere Temp. $T_p$ & $T_p=4.9\times 10^8 m^{-0.032}$ K &
$m^{-0.032}$\\

5.~~~Photosphere redshift $z_p$ & $1+z_p=1840 m^{0.343}$ &
$m^{0.343}$\\

6.~~~GBHC Rotation Rate, units $10^2~Hz$ - (RL06 47) &
$\nu_2=0.89[L_{q,32}/m]^{0.763}/L_{c,36}~ \approx 0.6/m~Hz$ &
$m^{-
1}~$\\

7.~~~GBHC Quiescent Lum., units $10^{32}~erg/s$ - (RL06 45, 46) &
$L_{q,32}=1.17 [\nu_2 L_{c,36}]^{1.31} = 4.8\times 10^{-3}
\mu_{27}^{2.62} \nu_2^{5.24} m^{-0.31}$ erg/s &
$m$~~\\
8.~~~Co-rotation Radius - (RL06 40)& $R_c=7\times 10^6
[m/\nu_2^2]^{1/3}$ cm &
$m$~~\\
9.~~~Low State Luminosity at $R_c$, units $10^{36}~erg/s$ (RL06
41) & $L_{c,36} = 0.015
\mu_{27,\infty}^2 \nu_2^3 / m$~~erg/s & $m$\\

10.~~Magnetic Moment, units $10^{27}~G~cm^3$ - (RL06 41, 47)&
$\mu_{27,\infty} = 8.16[L_{c,36} m/\nu_2^3]^{1/2}~~~\mu=1.7\times
10^{28}m^{5/2}~~~G~cm^3$ &
$m^{5/2}~$\\

11.~~Disk Accr. Magnetosphere Radius - (RL06 38) & $r_m(disc)=
8\times 10^6 [\mu_{27,\infty}^4/(m\dot{m}_{15}^2)]^{1/7}$ cm &
$m$~~\\

12.~~Spherical Accr. Magnetosphere Radius & $r_m(sp)$ or axial
 $z_m(in)= 2.3\times 10^7 [\mu_{27,\infty}^4/(m\dot{m}_{15}^2)]^{1/7}$ cm &
$m$~~\\

13.~~Spher. Accr. Eq. Mag. Rad. Rotating Dipole (RTTL03) &
$r_m(out)=1.2\times 10^7[\mu_{27}^2/(\dot{m}_{15} \nu_2)]^{1/5}$ cm & $m$~~\\

14.~~Equator Poloid. Mag. Field - (RL06 41, 47,
$B_\perp=\mu_\infty/r^3$)& $B_{\infty,10} = 250m^{-3}(R_g/r)^3
[L_{c,36}/(m^5 \nu_2^3)]^{1/2}$ gauss
& $m^{-1/2}$\\

15.~~Low State Jet Radio Luminosity - (RL04 18, 19)&
$L_{radio,36}=10^{-6.64}m^{0.84}
L_{x,36}^{2/3}[1-(L_{x,36}/L_{c,36})^{1/3}]$ erg/s & $m^{3/2}$~ \\
\hline
\end{tabular}\\
\end{table*}

\section{\bf Sgr A*}
Perhaps the strongest claimed evidence for an event horizon in any
black hole candidate is the one made for Sgr A* (Broderick and
Narayan 2006, hereafter BN06) based on its low radiated flux in
the near infrared. The bolometric luminosity ($\sim
10^{36}~erg/s$) and x-ray luminosity ($\sim 2\times 10^{33}~
erg/s$) of Sgr A* are also far lower than expected (Baganoff et.
al., 2003) from the standard thin accretion disk model used for
x-ray binaries and quasars. The bolometric luminosity is only
about $2\times 10^{-9}$ of the Newtonian Eddington limit rate for
an object of $3.7\times 10^6 M_\odot$. Nevertheless, if the
bolometric luminosity originated from accretion to a hard surface,
with 100\% efficiency, it would require an accretion rate of no
less than $10^{15}~g/s$ ($2\times 10^{-11} M_\odot/yr$). BN06
considered surface thermal radiations from an object with radius
in the range $2R_g < R < 100R_g$, where $R_g = GM/c^2 \approx
5.5\times 10^{11}~cm$ for a mass of $\sim 3.7\times 10^6 M_\odot$.
Subject to three critical assumptions they showed that if
redshifted, hard surface thermal emissions at $3.8 \mu m$ were
produced from such an accretion rate the radiated flux would be
too high unless the source radius were larger than about $40 R_g$.
But since the $3.5~ mm$ emissions of the compact radio source
apparently originate within $\sim 10 - 20 R_g$ of the central
object (Shen et al. 2005, Bower et al. 2004), one would expect the
NIR to originate within the same region. It is possible that
future VLBI measurements will further constrain the size of the
emitting region. The smaller the region, the more severe the
constraint on thermal emissions from the mass accretion rate. If
confined to the Schwarzschild diameter, the accretion rate in
accord with the assumptions of BN06 would have to be less than
about $3 \times 10^{13}~g/s ~(5\times 10^{-13} M_\odot/yr)$.

The assumptions on which these calculations for Sgr A* were based
are as follows: (1) A hard surface, possibly highly redshifted, (
$2R_g < R < 100R_g$) assumed to exist within Sgr A*, must radiate
in equilibrium with the accreting matter; i.e., the energy
transported to this redshifted hard surface by the accreting
matter must be radiated immediately with some nonzero efficiency
and must escape. (2) In addition the redshifted hard surface
assumed to exist in Sgr A* must radiate thermally at a temperature
in equilibrium with the rate of energy accretion and (3) General
Relativity is an appropriate description of gravity external to
the surface. BN06 stated that the current NIR flux density
measurements already conclusively imply the existence of an event
horizon for Sgr A*. Their conclusion is premature, however,
because it does not rule out the MECO model. Their assumption of
the existence of a hard surface does not apply to the MECO.

The MECO is blanketed by an optically thick pair atmosphere that
offers little impediment to accreting baryonic matter. Only about
$10^{-10}$ of accreting particle energy can be absorbed by coulomb
collisions with electrons or positrons in the pair atmosphere.
Collisions with photons in the pair atmosphere are somewhat more
effective at absorbing energy. About $10^{-7}$ of plasma accretion
energy can be absorbed by collisions of accreting electrons with
photons, however, only about one in $10^{10}$ of these compton
enhanced photons can get out through the extremely small general
relativistic escape cone (see Appendix A - D). The pair atmosphere
is essentially a phase transition region. The temperature is
buffered near the pair production threshold of $\sim 6\times 10^9$
K. Adding energy creates more pairs, but doesn't raise the
temperature. As a result of these considerations, it is apparent
that the MECO surface will remain quiescent until accretion
pressure would be a significant fraction of the local radiation
pressure. For the photosphere temperature of $3\times 10^8$ K
(Appendix C, Eq. 26) for a MECO model of Sgr A*, the local
radiation pressure would be $\sim 2\times 10^{19}$ erg cm$^{-3}$
(see Appendix D). At an accretion rate of $\sim 10^{17}$ g/s, (see
below) the maximum pressure accreting protons could contribute if
all of their momentum were stopped dead right at the photosphere
would only be about $10^{-6}$ of the radiation pressure already
present there. Since only about $10^{-7}$ of the momentum is
transferred to the entire pair atmosphere, it should be quite
clear that that the photosphere surface would remain quiescent for
this low accretion rate.

For a quiescent MECO, the observed radiated flux at frequency
$\nu_\infty $ and distance R ( $> 3R_g$) from the MECO is given by
(See Appendix A, Eq. 17)
\begin{equation}
F_{\nu_\infty} =\frac{2\pi h
\nu_\infty^3}{c^2}\frac{1}{e^{(h\nu_\infty/kT_\infty)} - 1}\frac{
27R_g^2}{R^2}
\end{equation}
With mass $m \approx 3.7 \times 10^6~ M_\odot$, Eq. 1, Table 1
gives a surface redshift of $z_s=1.1\times 10^{11}$. Eq. 2, Table
1 shows that the bolometric luminosity from a MECO of this mass
would be about $4.3\times 10^{33}$ erg/s. From Eq. 3, Table 1, its
distantly observed temperature would be $T_\infty=910~K$ with a
spectral peak at $3.2 \mu m$. For the most constraining (BN06) NIR
wavelength of $3.8 \mu m$, and the mass and $8~kpc$ distance of
Sgr A*, the quiescent MECO flux density given by Eq. 1 is 0.47
mJy, which lies below the observational upper limit by a factor of
three. Hence the general relativistic MECO model for Sgr A* shows
that a black hole with an event horizon is not required in order
to be consistent with the low luminosity observational constraints
of Sgr A*. But the constraint might also apply to flows that
produce external luminosity close enough to the central object.
For any model of the accretion process to satisfy the constraint,
whatever the nature of the central object, it is necessary to
limit the external accretion rate that gets within a few $R_g$ of
the object.

\section{Origins of Observed Radiations}
The multiwavelength spectrum of Sgr A* shows (e.g. An et al. 2005)
a relatively flat radio spectrum with a flux density dropping
steeply from a few $Jy$ at a few hundred $GHz$ to a few $mJy$ in
the NIR. The flat radio spectrum has been attributed to compact
relativistic plasma within a few $R_g$ of Sgr A*. The similar flat
radio spectra for GBHC are thought to arise from a jet outflow
(Markoff, Falcke \& Fender 2001, Falcke, K\"{o}rding \& Markoff
2003). It has been suggested that there could be a small jet
outflow from Sgr A* (Falcke \& Markoff 2000, Yuan, Markoff \&
Falcke 2002) but this is certainly not yet a consensus view. In
GBHC systems low state GBHC jets have been resolved (Stirling et
al. 2001) and studied over a wide range of GBHC luminosity
variation (Corbel et al. 2000, 2003). It has further been shown
(RL04, Heinz \& Sunyaev 2003) that the radio spectra are
consistent with mass scale invariant jets. Whether all (e.g.,
Heinz \& Sunyaev 2003), or only part (RL04), of the low state
x-ray emissions of GBHC originate in the base of the jet, it is
clear that the base of a jet can contribute.

The quiescent radiation from a MECO model for Sgr A* probably
would not originate in a jet outflow from an accretion disk.
Although more luminous AGN, modeled as MECO, are largely scaled up
versions of disk accreting GBHC, there are differences in
quiescence. In true quiescence for the MECO-GBHC the inner disk
radius lies beyond the light cylinder. But $T_\infty \approx
10^5~K$ for a central MECO-GBHC is much higher than for an AGN.
Even for a faint quiescent MECO-GBHC there would be a thermal
radiation flux capable of ionizing and ablating the inner regions
of an accretion disk out to $\sim 5\times 10^{3}R_g$. Ablated
material at low accretion rate would fall in and then be swept out
by the rotating magnetic field. This would produce the stochastic
power-law soft x-ray emissions in the MECO-GBHC quiescent state.
For Sgr A*, there would be only the cooler NIR radiation and
insufficient luminosity to ablate an inner disk with radius beyond
its light cylinder, hence nothing to keep the inner disk further
away. If the luminosity of the true quiescent state for Sgr A*
would correspond to a disk with inner radius at the light
cylinder, it would be at least $L_{q,max} = (2.7\times 10^{30}
erg/s) \mu_{27}^2 \nu_2^{9/2} m^{1/2} \sim 3 \times 10^{38}~erg/s$
(see RL06 Eq. 43 and Table 1 for magnetic moment and spin). But
since the luminosity of Sgr A* is well below this level, we can
conclude that its luminosity does not arise from a conventional
optically thick, geometrically thin accretion disk that extends in
to the light cylinder. This leaves a Bondi accretion flow as the
likely spectral source for a MECO model for Sgr A*.

\subsection{Bondi Accretion and Magnetic Propeller Effects}
It is expected that Sgr A* would accrete plasma in its vicinity
via a Bondi capture process. Based on plasma conditions within the
central parsec of the galaxy, an accretion rate of
$\dot{m}=3\times 10^{-6}M_\odot/yr=2\times 10^{20}~g/s$ and sound
speed of $c_{s,\infty}= 550~ km/s$ have been estimated (Baganoff
et al. 2003). The corresponding Bondi radius is $R_B=GM/c_\infty^2
= 1.6 \times 10^{17}~cm$. The expected accretion rate creates an
interesting conundrum. Even without any surface contributions an
accretion flow this large should create far more luminosity than
is observed even if it flowed into a central black hole. As shown
by Agol (2000), the strong polarization in the radio spectrum of
Sgr A* would further constrain the rate of accretion of a
magnetically equipartition plasma to be less than about $10^{-9}
M_\odot ~/yr$.

It is well known that stellar magnetic fields can severely inhibit
accretion to stellar surfaces (e.g., Toropina et al. 2003).
``Magnetic propeller effects" associated with stellar rotation
(Romanova et al. 2003, hereafter RTTL03, 2005) can cause
additional reductions. Revealing animations
of these processes can be viewed at\\
{\bf http://astrosun2.astro.cornell.edu/us-rus/}$~~.~~ $Results
displayed in Figure 6 of RTTL03 for the ``propeller flow regime"
show that there is a converging dense accretion flow to the
magnetic poles and a very low density, high speed toroidal
equatorial outflow. Significant accretion density variations occur
in the polar regions within a few times the object radius. The
conical flow into the polar regions provides a plausible compact
source of some luminosity from within a few $R_g$ of the central
object while the low density outflow would serve as the source of
most of the flat spectrum radio luminosity. The matter flowing in
on field lines that enter the polar regions can accrete to the
surface, but the bulk of the inflowing mass is ejected in the
magnetic propeller regime. RTTL03 demonstrated that as little as
2\% of the accreting material could reach the central star in
their simulations. An even smaller fraction should reach a more
compact central MECO with its much stronger magnetic field.

The propeller flow regime occurs if the magnetosphere radius lies
between the corotation and light cylinder radii. MECO co-rotation
and light cylinder radii are determined by the mass and spin
frequency for both GBHC and AGN (RL02, RL03, RL04, RL06), and
respectively, have the same range of values given by $\sim 50 -
100~ Rg$ and $500 - 1000~ Rg$. As noted by RTTL03, plasma within
the magnetosphere corotates with the central object, but only the
fraction that penetrates within the corotation radius can accrete
to the surface. Figure 7 of RTTL03 shows that beyond the
corotation radius, magnetic, centrifugal and pressure gradient
forces each exceed the gravitational force and combine to
accelerate the equatorial outflow to escape speed.

For the mass of Sgr A* ($3.7\times 10^6 M_\odot$), we expect a
spin frequency of $\nu = 1.6\times 10^{-5}$ Hz (Table 1, Eq. 6),
and a corotation radius of $r_c = 3.65 \times 10^{13}~ cm= 65R_g$
(Table 1, Eq. 8). The magnetic moment would be $\mu=4.5\times
10^{44}~ G cm^3$ (Table 1, Eq. 10). As shown by RTTL03, the Alfven
surface in a Bondi flow has a complex shape. The axial and
equatorial magnetospheric radii are unequal. RTTL03 showed good
agreement between the equatorial magnetosphere radius of their
simulation and the radius calculated from Eq. 13 Table 1, for
which the outflow is assumed to occur over about 30\% of the
$4\pi$ solid angle surrounding the dipole. The equatorial radius
is determined by equating the energy density of the magnetic field
to the kinetic energy density of matter thrown from the equatorial
magnetosphere. The equatorial magnetosphere radius for an
accretion rate of $3\times 10^{6} M_\odot/~~yr$ would be
$r_m(out)= 2.7\times 10^{14}~cm \sim 500 R_g$ (Table 1 Eq. 13),
which is not far inside the light cylinder radius of $3\times
10^{14}~cm$. It should be noted that the equatorial magnetic field
strength is still $\sim 20~-~40~~G$ in the region between
magnetosphere and corotation radii, but the kinetic energy density
of departing plasma is sufficient to drag the magnetic field lines
along with the outflow. In the outflow regions nearer the
corotation radius it can be seen in Figure 1 that much of the flow
is also loaded onto outward trending portions of the magnetic
field lines. The polar magnetosphere radius is determined by the
same balance of kinetic energy density and magnetic field energy
density along the dipole axis. Since the incoming flow merely
attaches to field lines having essentially the same direction, no
axial magnetospheric shock is expected. The polar axial
magnetosphere radius would be $z_m(in)=10^{15}~ cm \sim 1800 R_g$
(Table 1, Eq. 12). With the corotation radius well within the
magnetosphere radius a MECO model for Sgr A* would be in a strong
propeller regime.

\begin{figure}
\epsfig{figure=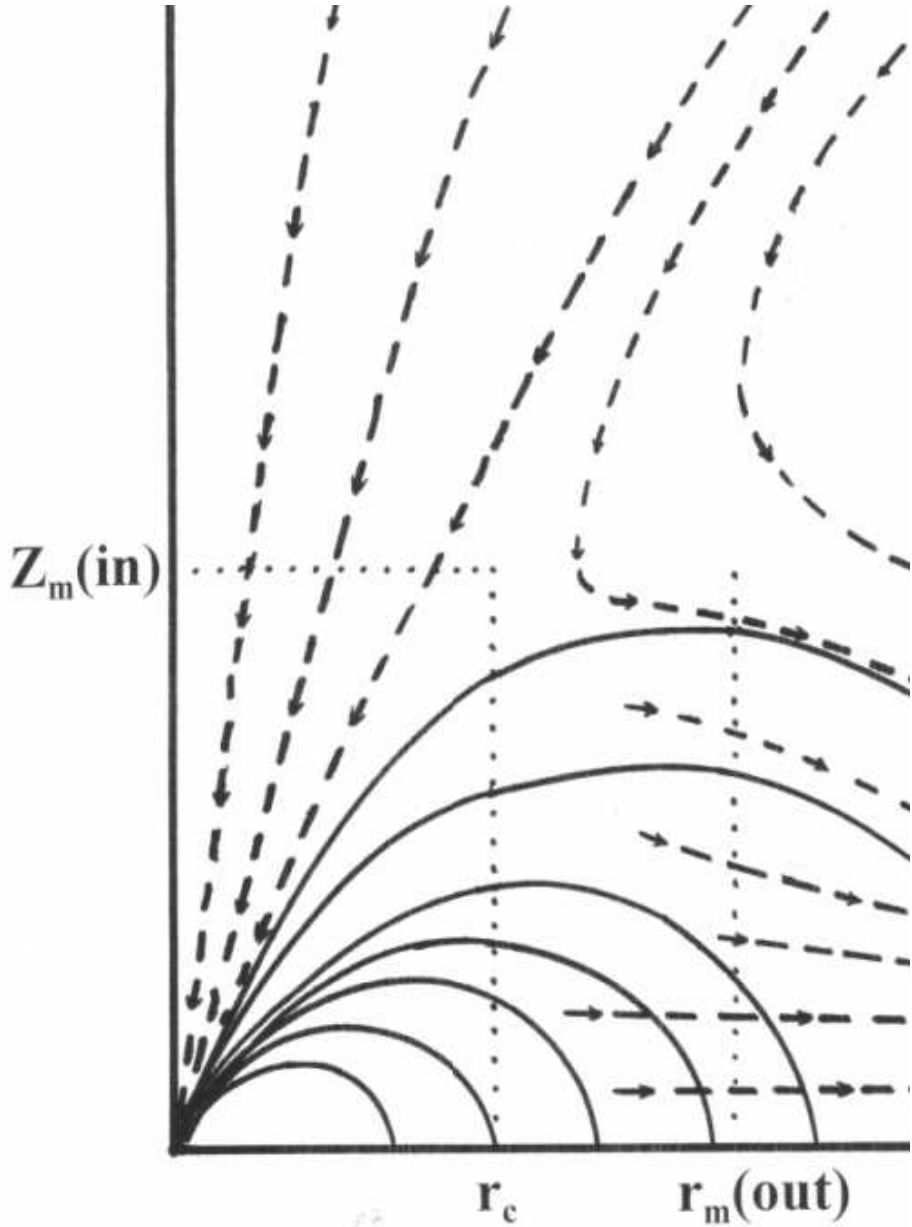,angle=0,width=15cm}\caption {Schematic
diagram of accretion flow into a rotating magnetic dipole. Solid
lines are magnetic field lines, dashed lines with arrows indicate
flow lines. Dotted lines mark corotation and magnetosphere radii.
The part of the flow that reaches the axial Alfven surface at
$z_m(in)$ and continues to the MECO surface is approximately
conical.}
\end{figure}

For a plasma flow to reach a rotating central dipole aligned along
the z-axis, it must enter inside the corotation radius. This can
occur easily for plasma flowing into the polar regions of the
magnetosphere. This part of the flow takes place in a spherically
symmetric conical pattern while the pattern elsewhere is a
circulating mix of inflow and outflow. A schematic diagram of the
flow is shown in Figure 1. The part of the flow that can reach the
central dipole enters at the top and bottom of a cylindrical
volume whose flat circular ends have a radius equal to the
corotation radius and whose height extends the axial magnetosphere
distance, $z_m(in)$ above and below the equatorial plane. The
parts of the flow that do not penetrate the corotation radius at
$z_m(in)$ are eventually ejected in the low density equatorial
outflow. The fraction of a Bondi flow that can reach the central
dipole is just the fraction of $4\pi$ steradians that is subtended
by the circle at distance $z_m(in)$. This fraction is $f=2 \pi
r_c^2/(4 \pi z_m(in)^2) = r_c^2/(2 z_m(in)^2)$. The factor of two
in the first numerator is for plasma entering both poles
\footnote{RTTL03 provided scalable relations for the fraction of
Bondi accretion that could reach the central magnetic dipole,
however, their simulations used a central object radius of
$R_*=0.0044R_B \approx 1300R_g$, which is too large for
application to the more compact magnetic MECO. For simulation
conditions that would be applicable to Sgr A*, the ratio of
corotation radius to Bondi radius would need to be $20\times$
smaller than the smallest ratio of RTTL03. The ratio of magnetic
field at the corotation radius to the field at the magnetosphere
radius would need to be about $2200\times$ larger than the largest
of RTTL03. In view of the comparatively smaller corotation radius
and larger magnetic field there we can expect the fraction of the
Bondi accretion rate that can reach the central object to be much
smaller than the minimal value of $f \sim 0.02$ found in RTTL03.}.
For $r_c=3.65\times 10^{13}~ cm$ and $z_m(in)=10^{15}~cm$, we find
$f=6.7\times 10^{-4}$. For a Bondi rate of $3\times 10^{-6}
M_\odot/~yr$ only $\sim 2\times 10^{-9} M_\odot/ ~yr=1.3\times
10^{17}~g/s$ would reach the central MECO\footnote{While this low
rate of flow reaching inside the corotation radius would satisfy
the proposed constraint imposed by observations of linear
polarization (Agol 2000), it is probably irrelevant because the
constraint entailed the assumption that the magnetic field was, at
most, an equipartition field generated within the flow. The
magnetic field strengths shown in Table 2 are so much larger that
they would produce strong polarization even for much higher
accretion rates.}.

Since plasma in the outflow cannot have gotten closer than the
corotation distance to the central MECO, we do not expect it to
contribute to the high frequency NIR spectral components observed
for Sgr A*. As described below, these, SSC x-rays and some thermal
brehmsstrahlung should be generated in the conical polar inflow.
The expanding equatorial outflow would be expected to produce flat
spectrum radio emissions similar to those produced by jets.
Although some radio emissions would also be produced within the
inflow region, the larger amount of outflowing plasma would
dominate the radio emissions. Other than the quiescent thermal
brehmsstrahlung contributions discussed below, most of the Sgr A*
spectrum consists of cyclotron/synchrotron radiations. Of course
it is necessary to know the distributions of magnetic field
strengths, plasma densities and temperatures before accurate
predictions of emission rates associated with the above processes
can be calculated. Global density, electron temperature and
magnetic field patterns for magnetic propeller flows similar to
those of RTTL03 will be needed in the future. Nonetheless we will
show in this paper that analytical methods can be used to produce
quantitative and qualitative predictions about the observed
spectral characteristics that provide a very plausible case for
the existence of a highly red-shifted supermassive MECO in the
center of Sgr A*.

As described in Appendix E, we have solved the energy equation for
spherical flow for application to the axial cones. The
characteristics in this part of the flow are determined primarily
by conditions at the Bondi radius. This solution clearly does not
apply to the more complex flow pattern outside the axial cones.
Flow speeds, ion densities and temperatures for the axial flows
are shown in Table 2. For a monatomic gas, the flow never becomes
supersonic, though it closely approaches sonic speed everywhere in
the polar flow within the magnetosphere. The sonic speed is half
what the free-fall speed would be. The general character of the
flow is such that ion density varies as $n \propto r^{-3/2}$, ion
temperature as $T \propto r^{-1}$, flow speed as $v \propto
r^{-1/2}$ and electron temperature as $T_e \propto r^{-1/2}$. As
described and calculated in Appendix E, the electrons gain the
bulk of their energy from collisions with ions but do not
thermally equilibrate with them. Electron temperatures, axial
magnetic field strength, cyclotron frequencies and Larmor radii
are shown in Table 2. The Larmor radii are much larger than the
mean particle spacing, $n^{-1/3}$ until $r \sim 30R_g$ and the
flow is optically thick until then. The flow is Compton thin
throughout.

For a number of reasons, we have limited calculations in Table 2
to $r \geq 3R_g$. First, the energy equation ceases to apply as
ion speeds become relativistic and it becomes necessary to
distinguish between coordinate speed and physical three-speed in
strong gravity. These distinctions do not change the qualitative
features obtained by using the energy equation below $30R_g$,
which is a little beyond its range of exact applicability. Second,
the luminosity generated near $3R_g$ is so refracted
gravitationally that it would appear to come from a larger region.
Third, the classical dipole expression for the magnetic field
begins to need modification as gravitational redshift becomes
significant inside $3R_g$ (Appendix B) . Lastly, the gravitational
redshift and Lorentz factors reduce the distantly observed
cyclotron radiation frequencies by more than the amount that the
relativistically enhanced magnetic field increases them. A
limiting frequency of a few times $10^{14}~ Hz$ is approached near
$3R_g$. In effect, the inflow synchrotron spectrum cuts off below
the NIR anyway. This interesting limiting frequency is set by the
the magnitude of the MECO surface magnetic field\footnote{A much
weaker magnetic field at $3R_g$ might not produce cyclotron
frequencies as high as the NIR. On the other hand, a much stronger
field might generate soft x-rays in the axial inflow.}. The
magnetic field is constrained to be no larger than would produce
bound electron-positron pairs on the MECO baryon surface. For this
reason it follows that strength of the MECO magnetic moment is not
a free parameter, but rather is a result of a quantum
electrodynamic stability constraint on the highly red shifted,
Eddington limited collapse process of the MECO suface described by
the Einstein-Maxwell equations (see RL06 and Appendix B).

\subsection{Spectral Characteristics of the Bondi Flow}
The plasma outside the magnetosphere can weakly radiate via
thermal brehmsstrahlung and synchrotron processes, but the
electrons are generally too cool to produce much luminosity.
Inside the magnetosphere strong, ordered magnetic fields exist and
cyclotron emission would be dominant until the accreting electrons
become mildly relativistic about at the corotation distance. As
shown in Table 2, different cyclotron fundamental frequencies are
generated in the polar inflow and the equatorial outflow, but both
inflow and outflow can contribute throughout most of the radio
frequency range to $\sim 10^{12}~~ Hz$. The reason for this is
that the mean particle spacing, $n^{-1/3}$ is much smaller than
the Larmor radius except in parts of the inflow that get inside
$\sim 30R_g$. The mean time between collisions, roughly the mean
particle spacing divided by the mean electron speed, is $\sim
10^{-12}~~s$ until a distance of $\sim 30R_g$ is reached. Hence
collision broadening will spread the cyclotron/synchrotron
frequency range to about $10^{12}~~Hz$ in both inflow and outflow.

For the MECO magnetic moment of $4.5\times 10^{44}~ G~~cm^3$, the
strongest magnetic field that electrons in the outflow could
encounter would be $\sim 9200~ G$ at the corotation distance on
the equatorial plane and the highest cyclotron frequency that they
would generate would be $26~ GHz$, however, as explained, the
spectrum produced by electrons that eventually depart in the
outflow would be collision broadened to as much as $10^{12}~~Hz$.
As the outflow continues outward it rapidly becomes less dense,
its emissions less collision broadened and generally of lower
frequency. Since the lowest cyclotron frequency generated within
the inflow and inside the corotation radius would be the $51~~GHz$
shown in Table 2, we can say that any lower frequencies would most
likely originate in the much larger mass that eventually departs
in the equatorial outflow. This picture is consistent with the
observation of a flare at $23~~GHz$ occurring later than the flare
at $43~ GHz$ (Yusef-Zadeh et al. 2008) and thus indicating an
outflow with lower frequencies produced further out in the
outflow. It would also not be surprising if the major contribution
to the earlier corresponding flare observed at $353~~GHz$ ($850
\mu~m$) also originated in the outflow, but closer to the
corotation boundary distance.

Since the electrons in the outflow cannot get within $30R_g$, they
can't contribute to frequencies above $\sim 10^{12}~~Hz$. Thus we
can say that the $10^{12}~~Hz$ to NIR range would be generated
entirely within the conical inflow. In addition, above
$10^{12}~~Hz$, the Larmor radius becomes smaller than the mean
particle spacing. The flow thus becomes optically thin inside
$25-30R_g$, with fundamental cyclotron frequencies dominating the
spectrum without significant collision broadening. Hence we see
that the morphology of the plasma polar inflow-equatorial outflow,
associated with the existence of an intrinsic MECO dipole magnetic
field in the center of Sgr A*, presents us with a natural way to
associate particular frequencies with different positions on the
axis ($\nu \propto B \propto z^{-1/3}$). What is even more
interesting is the fact that in context of Bondi accretion onto
Sgr A*, the differing spectral ranges of the equatorial outflow
and the polar inflow predicted by the MECO model for Sgr A* are in
agreement those required in the two spectral component model of
Sgr A* proposed by Agol (2000).

\subsection{Flares and Timing Considerations}
Recent measurements (Yusef-Zadeh et al. 2006, 2008) show that
x-ray and NIR flares preceded some corresponding flares observed
at radio frequencies. However, if NIR originates entirely in the
inflow, it would seem that flares in the radio would also be
produced earlier in the inflow at radio frequencies by plasma in
transit from greater distances. For example, a flare at
$10^{12}~Hz$ ($1000~GHz$) would be generated in the axial flow
field at $\sim 1.4\times 10^{13}~~cm=25R_g$ compared to a distance
of $3\times 10^{12}~~cm=5.5R_g$ for a NIR flare at $3\mu m$. At an
average flow speed of $\sim 7\times 10^9~~cm/s$ (Table 2), the NIR
flare should occur approximately 26 minutes later. Assuming that
the flares are caused by variations in the density of material in
the Bondi flow, we should actually expect two flares to occur in
the radio frequencies. A relatively weak early one should
originate in the inflow and a stronger one in the heavier outflow.
Differences in flow speed would likely cause them to occur at
different times. The inflow is straight down the magnetic field
lines at high speed while the plasma in the outflow would be
traveling more slowly and changing from inward to outward motion.
Its flare would most likely be delayed. As shown in Figure 1 of
Yusef-Zadeh et al. (2008) for observations of July 17, 2006 a weak
flare at $850 \mu m ~(353~~GHz)$ preceded an x-ray flare by about
half and hour, which was then followed about 1.4 hours later by
another much longer and brighter flare at $850 \mu m$. Another
observation showed (Yusef-Zadeh et al. 2006, Figure 1b) two NIR
flares, each of about one half hour duration and only one flare at
$850 \mu m$. If the latter is associated with the second NIR
flare, then it preceded the NIR by about 38 minutes. No $850 \mu
m$ measurements are shown beyond the occurrence of the second NIR
peak, so we don't know whether or not a larger radio flare
followed. Weak radio flares preceding NIR or X-ray flares would be
consistent with the MECO inflow. The larger second radio flare
occurring after the NIR/x-ray flares would also be expected. While
this picture associated with the MECO model is consistent with the
flare observations seen in Sgr A*, similar predictions could also
be made in terms of a black hole driven disk-jet model for Sgr A*
in which some radio frequencies are generated first in the flow
into a hot base of an outflowing jet where NIR and x-ray could be
produced. Presumably matter of increasing density could flow into
the base of a jet and produce some radio flaring before being
expelled. Then there could also be another set of radio flares
frequencies produced later further up a jet outflow.

Considering the weakness of the observed $850 \mu m$ flares that
preceded the observed NIR and X-ray flares, it is conceivable that
weak NIR/x-ray flares could be observed without noticeable earlier
flaring at radio frequencies in the Bondi inflow. Because of the
slower flows near the corotation radius in a magnetic propeller
outflow, a NIR/x-ray flare without much of a preceding radio flare
could still be produced before any radio flares, thus giving a
progression of flares that might all be thought to occur in an
outflow. Lastly, depending on the size and location of clumps of
matter in the incoming Bondi flow, it would be possible for
enhanced density to only occur in the outflow portion and produce
radio flaring without producing either x-ray or NIR flares.
Although these considerations make it seem plausible that we could
have combinations of NIR/x-ray or radio flares without always
having both, it seems likely that the strongest NIR/x-ray flares
would always be associated with both preceding and trailing radio
flares in the MECO model.

\begin{table*}
\tiny

\begin{center}
\caption{Bondi Flow Plasma, Magnetic and Spectral Parameters}
\end{center}
\begin{tabular}{lcccccccc} \hline
Axial or Radial Eq. Distance (cm)& $T_{9,ion}$ K & $T_{9,e}$~K& $v_z$ (cm/s) & $B_z$(G)& $\nu(polar)$&$\nu(out)$& $n(cm^{-3})$& $r_L$ (cm) \\
\hline

$1.6\times 10^{17}=R_B$& $0.022$ & $0.022$ & & & & & $26$ &\\
$10^{16}$ & $0.20$ & $0.09$ & $1.1\times 10^8$ & & & & $700$ &\\
$10^{15}=z_m(in)$ & $1.8$ & $0.29$ & $4.4\times 10^8$ & $0.9$ &
$2.5$ MHz & $1.25 MHz$ & $1.9\times 10^4$& $747$\\
$10^{14}\approx r_m(out)$ (equatorial) & $18$ & $0.93$ &
$1.5\times 10^9$ &
$900$ & $2.5$ GHz & $1.25$ GHz & $6.1\times 10^5$ & $1.4$ \\
$3.65\times 10^{13}=r_c$ & $49$ & $1.5$ & $2.5\times 10^9$ &
$1.8\times 10^4$ & $51$ GHz & $26$ GHz & $2.8\times 10^6$ &
$0.08$\\
$1.65\times 10^{13}=30R_g$ & $109$ & $2.3$ & $3.8\times 10^9$ &
$2\times 10^5$ & $560$ GHz & & $9.1\times 10^6$ & $0.01$\\
$10^{13}$ & $180$ & $2.9$ & $4.9\times 10^9$ & $8.9\times 10^5$ &
$2.5\times 10^{12}$ Hz &  & $1.9\times 10^7$ & $2.7\times 10^{-3}$\\
$1.65\times 10^{12}=3R_g$ & $1100$ & $7.2$ & $1.2\times 10^{10}$ &
$2\times 10^8$ &  $5.6\times 10^{14}$ Hz & & $2.9\times 10^8$ & $1.3\times 10^{-5}$ \\

\hline
\end{tabular}\\
\end{table*}

\subsection{Luminosities}
As shown in Appendix E, the luminosity produced in the axial
inflows to distance $z$ from the MECO is (Eq. 32)
\begin{equation}
L=4.2\times 10^{41} z^{-1/2}~~~ erg/s
\end{equation}
For $z=3R_g=1.65\times 10^{12}~cm$, this yields $L=3.3\times
10^{35}~ erg/s$. To $30R_g$ and $560~ GHZ$, the luminosity in the
polar flow would be $1.0 \times 10^{35}~ erg/sec$. For comparison,
the spectrum reported by An et al. (2005) shows a luminosity of
$1.7\times 10^{35}~ erg/s$ to a frequency of 560 GHz. The
discrepancy between these results occurs primarily in the lower
frequencies. The flux at $10^{12}~~Hz$ consistent with Eq. 2 is
about one order of magnitude below the observed radio spectral
trend. This should be expected because most of the radio spectrum
would be generated in the outflow, but Eq.2 should be fairly
accurate in the NIR which is produced only in the inflow.

At frequencies above about $10^{12}~Hz$, the axial flow is
optically thin and dominated by cyclotron fundamental frequencies
that can be associated with position in the dipole magnetic field;
i.e., $\nu \propto B \propto z^{-3}$. Thus inside $30 R_g$, it is
also shown in Appendix E that luminosity varies with frequency
along the axial flows such that (Appendix E Eq. 33)
\begin{equation}
dL =2\times 10^{32} \nu^{-5/6} d\nu~~~~~ erg/s
\end{equation}
Thus the spectral index in the optically thin IR/NIR would be
$-5/6$ for the conical inflow model. The average of four
measurements reported in BN06 in the NIR wavelength range from
$1.6 - 4.8 \mu m$ is $4.7~ mJy$. Using a distance to Sgr A* of $8~
kpc$, the average flux calculated here from Eq. 3 for the same
four wavelengths is $5.5~ mJy$. This result and the variability
shown by measurements at the position of Sgr A* strongly suggests
that the NIR spectrum originates in a variable accretion flow that
is consistent with the MECO model. While the flux calculated here
is somewhat larger than the lowest flux used in BN06 to constrain
an accretion rate to an assumed hard surface, it should be
remembered that variability in the accretion flow would produce
opportunities to observe both higher and lower luminosities. It
should also be noted that the calculated luminosity is sensitive
to the electron temperature, which is poorly constrained. On
balance, the agreement between MECO calculated and averaged
observed luminosities seems reasonable.

In addition to sensitivity to the electron temperature, the
luminosity generated in the inflow depends on the MECO spin rate
which then determines the size of the corotation radius ($\propto
\nu_s^{-2/3}$) and hence the fraction of the Bondi flow that can
reach the central MECO. We should expect the spin rate to be
influenced by angular momentum transported to the MECO by its
accretion environment. Nevertheless, the similar corotation radii
($50 - 100 R_g$) of disk accreting GBHC and AGN are necessary for
consistency with their mass scale independent radio cutoffs at
$\sim 0.02$ of Eddington limit luminosity. Given this mass scale
invariant property, it is not surprising that they seem to show
little variability in spins that scale as $\nu_s \propto m^{-1}$.
While we must admit that it is fortuitous that a similarly scaled
spin provides reasonable results for the MECO model of Sgr A*, it
also suggests that it may have had an accretion history similar to
that of other AGN.

For an optically thick slice in the inflow at frequencies below
$560 ~GHz$, the cyclotron spectrum would produce luminosity from
an axial region of thickness $dz$ that would be proportional to
$2\nu^2 kT_e/c^2$ times lateral surface area $2\pi r_c z/z_m dz$.
The correlation of frequencies with position, $z$, in the inflow
is surely weaker in the broad radio spectral band, but if we still
associate $z$ with frequency as $z \propto \nu^{-1/3}$, the
luminosity from the band would then be proportional to
$\nu^{2-1/3-4/3} d\nu$, hence the spectral index for the optically
thick band between $51~ GHz$ and $560~ GHz$ would be $1/3$,
compared to the reported $0.43$ (An et al. 2005). On the other
hand, if viewed looking parallel to the equatorial plane, the
luminosity in the outflow could be calculated in the same way in a
series of slices of thickness $dz$ with outflow radius
proportional to $\nu^{-1/3}$ ($\propto B^{-1/3}$) to obtain the
same spectral index. Both radio and NIR spectral indexes are
sensitive to the shape of the conical inflow pattern. Dipole
magnetic field lines have some curvature that would slightly widen
the cone at the top. If the ``cone" radius were proportional to
$z^b$ with $b > 1$, it would increase the overall luminosity and
spectral index calculated for the NIR and decrease the spectral
index in the optically thick region below the spectral peak. This
suggests that some refinements of our model might be expected to
produce closer agreement with observations but that is a task for
another time.

Electrons become mildly relativistic with Lorentz factors of
$\gamma_e \sim 2 - 3$ for small $z$. Some x-ray emission would
arise from Compton scattering within $\sim 30 R_g$ in the polar
inflow. Electron densities are also large enough in this region to
produce some thermal brehmsstrahlung. The calculation of
synchrotron self-compton (SSC) contribution can be started by
differentiation of Eq. 2 to obtain the luminosity contribution
from axial thickness $dz$ as $|dL|=2.2\times 10^{41} z^{-3/2} dz =
(dN_\nu/dt)E_\nu$, where $dN_\nu/dt$ is the rate of production of
synchrotron photons of average energy $E_\nu$ within $dz$. In
passing through distance $r=zr_c/z_m$, on average, these photons
will experience $n_e \sigma_T r$ collisions, where $n_e$ is the
electron density and $\sigma_T$ the Thompson cross-section. The
average energy gained per collision would be $2(\gamma_e+1)E_\nu
(kT_e/m_ec^2)$. Substituting for $n_e$, $T_e$ and $r$ as a
function of $z$, and integrating over $z$, gives the x-ray
self-synchrotron luminosity as
\begin{equation}
L_{SSC} = 3.8\times 10^{49}z^{-3/2} ~~~~~ erg/s
\end{equation}
which provides about $2\times 10^{31}~ erg/s$ to $3R_g$.

For a gaunt factor of $\sim 5$ for small z, thermal
brehmsstrahlung contributions can be calculated similarly. The
emissivity obtained is $6.2\times 10^9 z^{-2.75}~~ erg~ s^{-1}~
cm^{-3}~ Hz^{-1}$. Integration over the axial cone volume and all
frequencies yields a thermal (primarily x-ray) luminosity of
\begin{equation}
L_t = 3\times 10^{34} z^{-1/4} ~~~~~ erg/s
\end{equation}
This provides another $\sim 3\times 10^{31}~~erg/s$ to $3R_g$.

Though both calculated x-ray luminosities might be increased
somewhat by considering a flared ``cone", their combined
contributions should still fall well below the $2.4\times
10^{33}~~erg/s$ observed in the 0.5 - 7 keV band (Baganoff et al.
2003). Nevertheless, both depend on the square of electron density
which could enhance their contributions to the luminosity in
flares relative to synchrotron luminosity. These x-ray luminosity
variations in flares would be strongly correlated with the NIR
synchrotron luminosity variations and without time delays since
both originate in the same population of electrons.

A substantial fraction of the quiescent x-ray luminosity appears
to come from a spatially extended source (Baganoff et al. 2003).
There is a considerable volume in the MECO magnetosphere in which
temperatures and densities would be high enough to produce thermal
x-rays. Within the rough bounds of the magnetosphere outside the
corotation radius there is a volume of $2 \pi r_m(out)^2 z_m(in)
\sim 3\times 10^{44}~~cm^3$. Using $10^{14}~~cm$ as an average
radius in the Bondi flow, we estimate the corresponding electron
temperature to be $10^9~~K$ and the electron density to be
$10^6~~cm^{-3}$. For these parameters, a gaunt factor of 3 and the
$1.6\times 10^{18}~~Hz$ bandwidth from 0.5 - 7 keV, we find an
average thermal brehmsstrahlung emission rate of $10^{-11}~~erg
s^{-1} cm^{-3}$. Multiplying by the magnetosphere volume, we
obtain $\sim 3\times 10^{33}~~erg/s$, which is reasonably close to
the observed quiescent x-ray luminosity of Sgr A*.

The axial inflows constitute such a small fraction of the Bondi
accretion rate that their contributions to the spectrum below $52~
GHZ$ are much smaller than the radiation from the much larger
outflow. We don't know how much the outflows might contribute in
harmonic frequencies above $26~ GHz$, but since the luminosity
calculated for the axial flow did not account for the observed
luminosity or flux to a frequency of $560~ GHz$, there is a
significant contribution from the outflow that has not been
considered. Accurately quantifying the radio luminosity will
require knowledge of the plasma and magnetic field distributions,
as previously mentioned. All that we can do here is set an upper
limit on what might be produced in the outflow. For an accretion
rate of $2\times 10^{20}~ g/s$, electrons would flow outward at a
rate of $1.2 \times 10^{44}~/s$. If they each had the energy they
could extract from the protons while reaching the corotation
radius, there would be about $2 \times 10^{37}~ erg/s$ available
for radiation. The actual radiation rate would probably be much
smaller as few of the electrons that would flow out would reach
such small distances from the MECO. In this regard, in reaching a
steady state, the magnetic propeller eventually must drive the
outflow beyond the Bondi radius, cutting off a large fraction of
the flow into the Bondi sphere and eventually producing a large
torus where the outflow stops beyond the Bondi
radius\footnote{Note that the simulations of RTTL03 never reached
this steady state. Cutting off a substantial fraction of the
accretion flow could have a significant effect on the inflow rate
and thus affect the luminosity generated in the outflow beyond the
corotation radius, but it probably would not greatly affect our
estimated luminosity in the polar flow, which depends primarily on
the spherical flow pattern in the polar regions and the asymptotic
values of density, $n_\infty$ and temperature $T_\infty$ at the
Bondi radius.}. A luminosity of a few times $10^{36}~ erg/s$ would
seem to be entirely reasonable.

\begin{figure}
\epsfig{figure=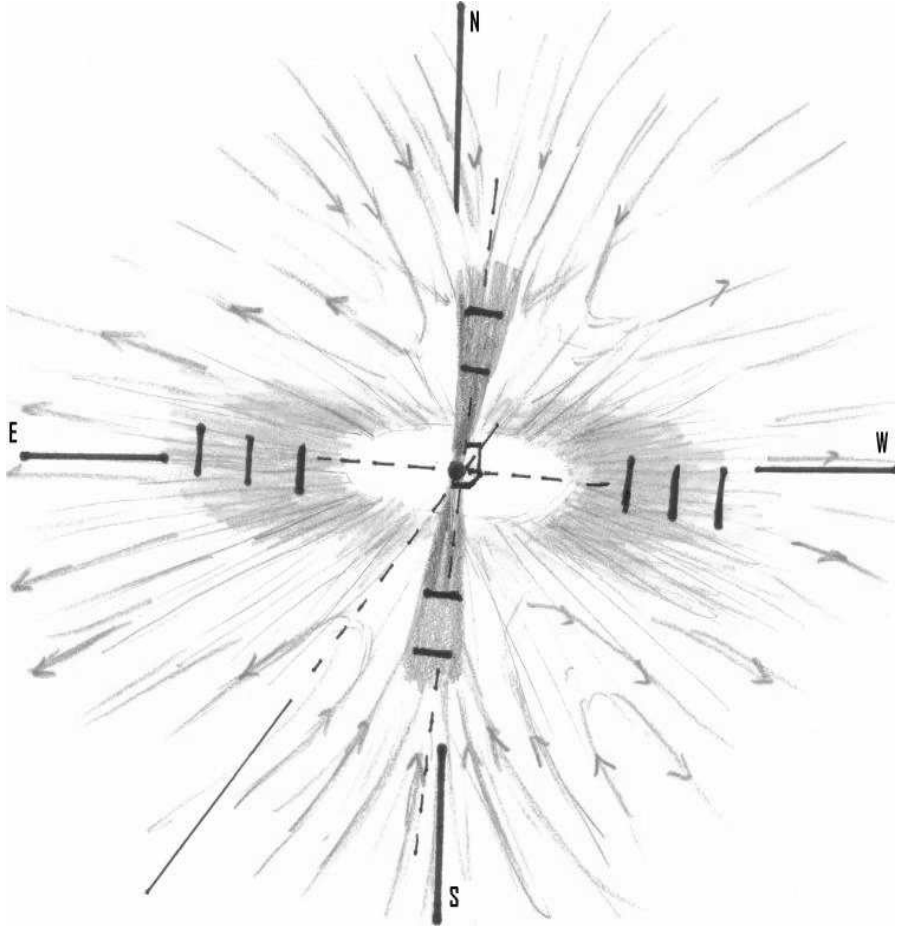,angle=0,width=15cm}\caption {Sketch
of Bondi inflow/outflow pattern within the magnetosphere for MECO
model of Sgr A*. Most of the radio spectrum to $10^{12}~Hz$ is
generated in plasma that departs in an equatorial disk-like
outflow. The disk outflows are seen limb brightened, generally W-E
and E-W as seen against the sky plane. NIR would be produced in
the generally N-S and S-N axial inflows. Polarization directions
marked with dark bars are essentially perpendicular to magnetic
field lines. Quiescent x-ray luminosity would be produced
throughout the magnetosphere volume. X-ray SSC and thermal
brehmsstrahlung radiation would be produced in flares in the axial
inflow NIR region. The luminosity of the central MECO, even at its
peak luminosity at $3.2\mu m$, is below observational limits.}
\end{figure}

\subsection{Polarization}
The MECO-Bondi flow pattern has two major regions of strong,
ordered magnetic fields, with the strongest fields along the polar
axis. With the NIR generated within the the polar inflow, its
linear polarization direction would be essentially perpendicular
to the axis. The equatorial outflow would produce magnetic field
lines stretched generally radially along the outflow. These field
lines would be perpendicular to the magnetic axis and there are
several possibilities for what might be observed, depending on the
orientation of the magnetic axis and equatorial plane relative to
an observer. Yusef-Zadeh et al. (2006) showed some evidence of
outflow from Sgr A*. The apparent flow was in a generally WSW
direction from Sgr A*. Proper motion studies have provided
confirmation (Muzic et al., 2007) of a general, uncollimated
outflow. GRAVITY, a recently proposed high resolution
interferometric imaging system (Eisenhauer 2008), should be
capable of revealing the nature of the plasma flows in the
vicinity of Sgr A* in unprecedented detail. If what has been
observed at radio frequencies would be a view into the MECO-Bondi
equatorial outflow, we would see the limb brightened regions of
the outflow extending WSW (and ENE) against the plane of the sky,
with magnetic field lines stretched out along the flow. Linear
polarizations from the outflow would then be generally
perpendicular to the radio luminosity and in the SSE-NNW
direction. A sketch of the way this might appear is shown in
Figure 2. The radio polarization has been observed at $230~GHz$
(Bower et al. 2003) and clearly displayed, aligned perpendicular
to the WSW-ENE direction of the limb brightened part of the
outflow. (See Fig. 2.1,
http://www.cfa.harvard.edu/sma/newsletter/smaNews\_21Dec2006.pdf).
Weaker polarization would be produced further out in the flow
where the field is weaker and fundamental cyclotron frequencies
would be lower.

In addition to the polarization direction at radio frequencies
matching expectations for an outflow, strong polarization has been
observed in the NIR with a direction roughly 90 degrees (ENE)
different from that seen at radio frequencies (Eckart et al.
2008). As noted, this is consistent with the polarization that
should be seen perpendicular to the field lines in the inflow if
viewed at high inclination to the inflow axis. A view with this
orientation would be a view fairly directly into the equatorial
outflow and directly down the magnetic field lines of the outflow.
Because of the dipole nature of the MECO magnetic field, a view
directly into the equatorial plane would have outgoing magnetic
field lines on one side of the the plane and ingoing field lines
on the other. Outflowing electrons would produce left circular
polarization from outflow on one side of the equatorial plane and
right circular polarization from the other. Since it is unlikely
that the thickness of the outflow would be resolved, there would
be a net circular polarization only if viewed at some inclination
to the outflow equatorial plane. This could occur while still
having a large inclination relative to the inflow axis. Electrons
spiraling outward on field lines directed toward us would produce
a (negative) left circular polarization. The circular polarization
($\sim 1\%$) of Sgr A* (Bower 2000) is unusually strong for an AGN
and approximately double the degree of linear polarization at the
same frequency. It seems unlikely that it would originate in a
depolarizing medium. While it exhibits short-term variability, it
has maintained a stable (negative) sense for over 25 years. In
this context it is important to note that the observation of
stable directions, with some variability in flares, for both
linear and circular polarizations are unique observable features
of the MECO model for Sgr A*. All of these properties are
determined by the strength of the intrinsic magnetic field of the
MECO within the center of Sgr A*.

\subsection{Image Appearances}
The observed polarizations are consistent with a view close to the
equatorial plane of the outflow and across the axial inflow as
shown in Figure 2. For this orientation we should see an
increasingly wide outflow zone at longer radio wavelengths. Inside
any surface of constant magnetic field strength cutting through
the outflow, we would expect optically thick cyclotron radiant
flux of $j_\nu = 2\nu^2 kT_e/c^2$ to be generated. The apparent
size of its limb brightened photosphere image in the equatorial
outflow would be roughly\footnote{The limb brightened region
viewed would not be a surface of constant magnetic field and would
not correspond to a size exactly proportional to $\lambda^2$
throughout.} proportional to $\nu^{-2} \propto \lambda^2$ and
ellipsoidal as has been observed (Bower et al. 2004). The major
axis of the ellipsoid is apparently aligned with the direction of
outflow (Muzic et al. 2007, Yusef-Zadeh et al. 2006, Bower et al.
2004) and perpendicular to the radio polarization direction (Bower
et al. 2003). These observations are consistent with our
prediction that the apparent size and ellipsoidal shape of the
radio images of Sgr A* are due to the equatorial outflow from a
MECO in the center of Sgr A*.

Scattering theory also predicts image sizes proportional to
$\lambda^2$. The image sizes (Bower et al. 2004, Shen et al. 2005)
have been interpreted only as scattering features superimposed on
a compact object even though the major axis of the observed
ellipsoids is twice the length of the minor axis. The galactic
center scattering screen is two to three orders of magnitude
greater than what is seen in NGC6334B, the next most scattered
source (Bower et al. 2004). A heavy screen would be expected for
the base of the MECO outflow where its highest radio frequencies
would be generated. The claimed detection of an intrinsic size of
$48R_g$ for Sgr A* rests on apparent deviations from the
scattering theory with its wavelength exponent of 2, but for the
cyclotron-synchrotron radiations considered here the exponent and
deviations at distances inside the corotation radius could arise
in a different way. Only much less luminous radio emissions should
be generated in the inflow at wavelengths below about 0.3 mm
($10^{12}~Hz$). Eq. 33 (Appendix E) gives an expected flux of
$0.25~ Jy$ for this wavelength and a distance of $8~kpc$ to Sgr
A*.

As shown above, the surface emissions from a MECO model for Sgr A*
would peak well below detection limits at $3.2 \mu m$. At
wavelengths differing by a factor of even two from the thermal
peak a MECO would be as dark as a black hole (see Appendix A - C).
Since we identify the $10^{12}~Hz$ to NIR spectrum as originating
in concentrated axial flows into the magnetic poles, we would
expect image sizes at optically thin wavelengths to no longer be
proportional to $\lambda^2$ and to be elongated along the inflow
axis rather than the equatorial plane. In the optically thin flow,
axial distance from the MECO should be correlated with frequency
as $z \propto \nu^{-1/3} \propto \lambda^{1/3}$. Hence on the
basis of the MECO model for Sgr A* we are led to the prediction
that, if viewed at high inclination to the polar axis, there would
be two axial lobes, beginning at about $25R_g$ and extending ever
closer to a central dark source at shorter wavelengths. Depending
on the inclination there might be significant Doppler boosting in
the lobe with flow components directed more toward us, however at
these radial distances there would not be a uniform background of
sub-mm to NIR radiation against which a very dark shadow of the
MECO could be viewed. There would be such a background only for
wavelengths that originated inside $\sim 5R_g$ and were extremely
refracted gravitationally. These wavelengths would necessarily be
in the NIR.

The MECO-Bondi model presented here can be compared with the
radiatively inefficient accretion flow (RIAF) into a black hole
(Yuan, Quataert \& Narayan 2003). To produce the observed
polarization of the optically thick radio emission, the internally
generated magnetic field of the RIAF would need to be oriented
generally E-W. This might occur in two different ways. (a) If the
RIAF magnetic field responsible for the polarization were toroidal
within the thick RIAF disk, the RIAF disk would need to be seen
nearly edge-on in order to have a consistent magnetic field
direction as seen from our side. There would only be one direction
of flow observed, either E-W or W-E, in the radio. Accordingly, a
suitable radio interferometric array analogous to the proposed NIR
GRAVITY array (Eisenhauer et al. 2008) might be able to
distinguish a RIAF from a MECO outflow. But in the NIR, the RIAF
flow might show no preferential direction as observed with the
GRAVITY array because both near and far sides could be seen at
once if sufficiently optically thin. The NIR, which would be
produced closer to the central black hole is optically thin and
would be produced in part by non-thermal electrons. It is the
power-law distribution of non-thermal electron energies in the
optically thin part of the flow that would produce the orthogonal
polarization of the NIR (Agol 2000). (b) If the RIAF magnetic
field were poloidal; i.e., with field lines emerging perpendicular
to the disk, then the disk would need to be oriented generally
N-S, but again seen nearly edge on. Again the flow would appear to
be unidirectional. The orthogonal polarization of the NIR would
then arise in the same way as before. Thus with either
orientation, both polarizations would originate in the
geometrically thick disk of the RIAF and it should not look the
same as the two zone flow of the MECO-Bondi model shown in Figure
2.

It is clear that the accretion rate in a RIAF must be of order
$10^{-9} M_\odot ~/yr$, and far below the Bondi rate of $3\times
10^{-6} M_\odot~/yr$ in order to produce the observed strong
linear polarization within the flow. Exactly how the RIAF might
accomplish this is unknown. It has been suggested that the disk
evaporates in a wind or that a jet outflow occurs. For a jet
outflow to be consistent with observations (Yusef-Zadeh et al.
2006, Muzic et al. 2007, Bower 2003) the RIAF disk would need to
be aligned generally N-S with a generally E-W jet. Though this
would give more of the appearance of Figure 2, a jet would be an
actual collimated outflow rather than just a limb brightened
region of a toroidal outflow and the flow in the RIAF disk would
still be unidirectional rather than into the central region from
both above and below as in the flow into the poles of a MECO. The
GRAVITY array should be able to distinguish these possibilities.
Ironically, if inflows were observed to disappear into an unseen
dark object from both above and below, it could be mistaken for
the flow into a black hole event horizon, but some explanation
would then be required for an equatorial outflow which ought to
also be observed\footnote{ While the expected appearances of MECO
and RIAF models are sufficiently different to eventually be
clearly observed, there are some theoretical objections that can
be raised about the RIAF model. 1) The RIAF model is aesthetically
unappealing since it appears to be based on a collection of ad hoc
dynamic assumptions which have been added onto the original ADAF
model in order to allow it to be able to explain the Sgr A*
observations. 2) One has to assume a consistent long term angular
momentum within the Bondi inflow into Sgr A* in order to get the
flow to circularize and form an accretion disk. 3) Assuming that
such a Black Hole RIAF accretion disk can be formed by the Bondi
inflow, it is then required that most of this disk has to
``evaporate" and ultimately escape as a wind or be ejected in a
jet outflow. If the latter, the jet must be formed at considerable
distance, probably beyond $20-30R_g$, by means not presently
known. There would be too much luminosity generated by a
Kerr-metric ergospheric jet. 4) If the flow if hot enough to
produce the quiescent soft x-rays over an extended region, why
does it not produce more and harder luminosity in the RIAF part of
the flow 5) While it is clear that a source of non-thermal
electrons must exist within the black hole RIAF model there still
remains an unsolved problem about the source of their energy.
Clearly these non-thermal electrons cannot have been accelerated
by magnetic reconnection processes because if that were the case
then the black hole RIAF would be too bright in the NIR to fit the
Sgr A* observations. In spite of these problems, a black hole RIAF
can be used to account for most of the spectral distribution that
has been observed, provided that it would be allowed to produce
some radio emission in the disk as well as the jet. This is
necessary in order to account for the observed timing of NIR/x-ray
flares with both preceding and trailing radio flares.}.

\section{Summary}
We have shown that the MECO model accounts for the lack of
observed surface luminosity from Sgr A*. Although it is necessary
to know the global distributions of electron density and magnetic
fields before spatial and spectral energy distributions can be
accurately calculated, we have shown that a spectrum of
approximately the correct luminosity and spectral indexes would be
produced in the inflow-outflow zones of a Bondi accretion flow
into the magnetic field of a MECO. The equatorial outflow would
produce optically thick cyclotron radiation with positive spectral
index at frequencies below $\sim 1000~~GHz$. The axial inflow
would produce steeply declining (negative index) optically thin
NIR emissions as well as some correlated x-ray SSC and
brehmsstrahlung emissions that could be observed in flares caused
by high density clumps in the inflow. We have shown that timing of
flares in radio/NIR/x-ray bands is consistent with the MECO-Bondi
model, with some weak sub-mm flaring preceding the strongest NIR
and SSC x-ray flares which are then followed by stronger delayed
sub-mm and radio flare emissions.

The part of the Bondi flow that eventually departs in the
equatorial outflow would produce radio emissions, possibly to
frequencies as high as $\sim 1000~~GHz$, and including nearly
everything below $\sim 50~~GHz$. The bulk of the quiescent x-ray
luminosity would be thermally generated within the magnetosphere
in a mixed inflow/outflow pattern of $10^{14-15}~~cm$ size. We
have shown that the low bolometric luminosity of Sgr A* can be
reconciled with an expected Bondi accretion rate in a completely
natural way. The magnetic propeller mechanism is a robust, stable
physical mechanism for reducing the Bondi accretion rate to levels
compatible with the low luminosity of Sgr A*. The only parameters
that have been necessary for these calculations are the ion
density and sound speed at the Bondi radius, mass, magnetic moment
and spin. The first three of these have been taken from work
reported by others. The intrinsic magnetic moment of a MECO is an
inherent, mass dependent feature which is generated by the effects
of the quantum electrodynamic stablility conditions that are
required by the Einstein-Maxwell equations that describe the
highly red shifted Eddington limited MECO collapse process (RL06).
Its magnitude sets a natural high frequency limit for the
synchrotron emissions in the axial inflow. Our estimate of spin
has been taken from our previous work that has accounted for the
radio/x-ray luminosity correlations and spectral state switches
for AGN and GBHC (RL04) and the microlensed image of the quasar
Q0957+561 (Schild, Leiter \& Robertson 2006). In retrospect, it is
fortuitous that this choice has worked out well, but it suggests
that the accretion history of Sgr A* may be similar to that of
other AGN.

Since they do not possess event horizons, highly red shifted
general relativistic MECO contain intrinsic magnetic moments that
can interact with their environments. The intrinsic MECO magnetic
moment automatically produces the ordered magnetic fields
necessary to account for the observed strong linear polarization.
In the MECO-Bondi inflow-outflow model for Sgr A* described here,
there are regions of differing magnetic field strength and
orientation, strong density variations and both inflows and
outflows. We have shown that there are places of origin in this
flow for all of the spectral, spatial, polarization and timing
features that have so far been observed for Sgr A*. The patterns
of inflow and outflow differ from those expected of black hole
models and should be observable by the proposed GRAVITY array
(Eisenhauer et al. 2008). On the basis of the MECO-Bondi model for
Sgr A* we predict that high resolution images in radio frequencies
should be elongated in the limb brightened outflow zone, while
high resolution images in shorter wavelengths should be elongated
along an orthogonal polar axis. Since the emissions in these
shorter wavelengths are confined to the narrow axial inflow
region, there would be no uniform background to provide a
silhouette image of a dark MECO except for strong gravitational
refraction effects on NIR frequencies generated inside $\sim
5R_g$. Everything inside $\sim 25R_g$ would just be dark in the
radio frequencies.

The qualitative consistency of the MECO-Bondi model provides added
incentive for doing additional simulations which, in order to
succeed, will require substantial computational facilities and
expertise. Calculations of the global magnetic field and density
distributions will be necessary first steps before synchrotron
emissions in the outflow can be calculated. In order to add
spectral details, smaller calculation grids close to the central
dipole and consideration of light paths in strong field gravity
will be required and could be challenging, even for the original
US-Russia supercomputer collaboration that produced the work of
RTTL03. Calculations extending into the region near the photon
sphere ought to be able to provide spectral and image details that
could be compared with the increasingly high resolution images of
Sgr A*. Further corroboration of the MECO-Bondi model for Sgr A*
could be found if further observations eventually reveal NIR lobes
for accretion flow into the magnetic polar regions, though they
would likely be smeared into ellipsoids by gravitational
refraction. A pattern of expected polarizations of radiation
should also be computed. The MECO parameters for spin and magnetic
moment used here provide a place to start, but for best
comparisons these parameters, as well as the Bondi flow parameters
should be varied. Rather detailed calculations of what might be
seen for different axis orientations will also be necessary. It is
our hope that the present paper will provide motivation for the
additional work to be done.

\section{Appendix}
{\bf A. ECO Models}\\

An ``eternally collapsing object" ECO is a gravitationally compact
mass supported against gravity by internal radiation pressure
(Mitra 2006). In its outer layers of mass, a plasma with some
baryonic content is supported by a {\it net} outward flux of
momentum via radiation at the local Eddington limit $L_{Edd}$
given by
\begin{equation}
L_{Edd,s}=\frac{4 \pi G Mc (1+z_s)}{\kappa}
\end{equation}
Here $\kappa$ is the opacity of the plasma, subscript s refers to
the baryonic surface layer and $z_s$ is the gravitational redshift
at the surface. In General Relativity $z$ is given by
\begin{equation}
1+z=1/\sqrt{1-2GM/c^2R}
\end{equation}
For a hydrogen plasma, $\kappa=0.4$ cm$^2$ /g and
\begin{equation}
L_{Edd,s}=1.26\times 10^{38} m (1+z_s) ~~~~erg/s
\end{equation}
where $m=M/M_\odot $ is the mass in solar units.

Since the temperature at the baryon surface is beyond that of the
pair production threshold, there is a pair atmosphere further out
that remains opaque. The net outward momentum flux continues
onward, but diminished by two effects, time dilation of the rate
of photon flow and gravitational redshift of the photons. The
escaping luminosity at a location where the redshift is $z$ is
thus reduced by the ratio $(1+z)^2/(1+z_s)^2$, and the net outflow
of luminosity as radiation transits the pair atmosphere and beyond
is
\begin{equation}
L_{net~out}=\frac{4 \pi G Mc (1+z)^2}{\kappa (1+z_s) }
\end{equation}
Finally, as distantly observed where $z \rightarrow 0$, the
luminosity is
\begin{equation}
L_\infty=\frac{4 \pi G M c}{\kappa (1+z_s) }
\end{equation}
For hydrogen plasma opacity of 0.4 cm$^2$ g$^{-1}$, and a typical
GBHC mass of $7~M_\odot$ this equation yields $L_\infty=8.8 \times
10^{38}/(1+z_s)$ erg s$^{-1}$. But since the quiescent luminosity
of a GBHC must be less than about $10^{31}$ erg s$^{-1}$, we see
that it is necessary to have $z_s > 8.8 \times 10^7$. Even larger
redshifts are needed to satisfy the quiescent luminosity
constraints for AGN. This is extraordinary, to say the least, but
perhaps no more incredible than the $z = \infty$ of a black hole.

At the low luminosity of Eq. 10, the gravitational collapse is
characterized by an extremely long radiative lifetime, $\tau$
(Robertson \& Leiter 2003, Mitra 2006) given by:
\begin{equation}
\tau= \frac{\kappa c(1+z_s)}{4 \pi G}=4.5\times 10^8 (1+z_s)~~yr
\end{equation}
With the large redshifts that would be necessary for consistency
with quiescent luminosity levels of BHC, it is clear why such a
slowly collapsing object would be called an ``eternally collapsing
object" or ECO.

For $(1+z) > \sqrt{3}$, radiation is impeded by passage through a
small escape cone such that the fraction of radiation that could
escape if isotropically emitted at radius R would only be
\begin{equation}
f= 27\frac{R_g^2}{R^2(1+z)^2}
\end{equation}
For very large $z$, $R \approx 2R_g$ and $f=(27/4)(1+z)^{-2}$.

At the outskirts of the pair atmosphere of an ECO the photosphere
is reached. Here the temperature and density of pairs has dropped
to a level from which photons can depart without further
scattering from positrons or electrons. Nevertheless, the redshift
is still large enough that their escape cone is small and most
photons will not travel far before falling back through the
photosphere. If we let the photosphere temperature and redshift be
$T_p$ and $z_p$, respectively, the net escaping luminosity is
\begin{equation}
L=\frac{27R_g^2 4 \pi R^2 \sigma T_p^4}{R^2(1+z_p)^2}=\frac{4 \pi
G Mc(1+z_p)^2}{\kappa (1+z_s) }
\end{equation}
But in the radiation dominated region beyond the photosphere, the
temperature and redshift are related by
\begin{equation}
T_\infty=\frac{T}{1+z}
\end{equation}
where $T_\infty$ is the distantly observed radiation temperature.
Substituting into the previous equation, we obtain
\begin{equation}
L_\infty=(27) 4 \pi R_g^2 \sigma T_\infty^4=\frac{4 \pi G
Mc}{\kappa (1+z_s) }
\end{equation} and, for hydrogen plasma opacity
\begin{equation}
T_\infty = \frac{2.3\times 10^7}{[m(1+z_s)]^{1/4}}~~~~K
\end{equation}
The left equality of Eq. 15 can be written in terms of the
distantly observed spectral distribution, for which the radiant
flux density at distance R ($
> 3R_g$) and frequency $\nu_\infty $ would be
\begin{equation}
F_{\nu_\infty} =\frac{2\pi h
\nu_\infty^3}{c^2}\frac{1}{e^{(h\nu_\infty/kT_\infty)} - 1}\frac{
27R_g^2}{R^2}
\end{equation}

\noindent {\bf B. MECO}

As previously discussed (RL06), if one naively assumes that the
photon support for an ECO originates from purely thermal
processes, one quickly finds that the temperature in the baryon
surface layer would be orders of magnitude higher than the pair
production threshold. The compactness guarantees (Cavaliere \&
Morrison 1980) that photon-photon collisions would produce
numerous electron-positron pairs. Drift currents proportional to
${\bf g\times B}/B^2$ reactively generate extreme magnetic fields.
We assumed that the baryonic surface field reaches an equipartion
level with a surface magnetic field of $\sim 10^{20}$ G (Harding
2003, Zaumen, 1976) that is capable of creating bound
electron-positron pairs on the baryon surface. The interior
magnetic field in a stellar mass MECO-GBHC is about what would be
expected from flux compression during stellar collapse. At the
MECO surface radius ($R \approx 2R_g$), the ratio of tangential
field on the exterior surface to the tangential field just under
the MECO surface is given by (RL06)
\begin{equation}
B_{\theta,S^+}/B_{\theta,S^-}=(1+z_s)/(2ln(1+z_s))
\end{equation}
\begin{displaymath}~~~~~~~~~~ = 10^{20}~G/(B_{in}\sqrt{7M_\odot/M})
\end{displaymath}
We have previously taken $B_{in}=2.5\times 10^{13}$ gauss as
typical of the interior field that can be produced by flux
compression during stellar gravitational collapse. Using this
value the preceding equation has the solution.
\begin{equation}
1+z_s=5.67\times 10^7 m^{1/2}
\end{equation}
The magnetic moment of a MECO would be
\begin{equation}
\mu=1.7\times 10^{28} m^{5/2}~~~~~~G~ cm^3
\end{equation}
We have found that this magnetic moment, and the spin rates given
by Table 1, Eq. 6 give the MECO model a good correspondence with
observations of spectral state switches and the radio luminosities
of jets for both GBHC and AGN (RL02, RL03, RL04, RL06).

\medskip
\noindent{\bf C. The Photosphere}

The photosphere, as a last scattering surface, can be found from
the condition that (Kippenhahn \& Wiggert 1990)
\begin{equation}
\int_\infty^{R_p} n_\pm \sigma_T dl = 2/3
\end{equation} where $dl$ is an increment of proper length in the
pair atmosphere and $n_\pm$ is the combined number density of
electrons and positrons along the path. Landau \& Lifshitz (1958)
show that
\begin{equation}
n_\pm=\frac{8\pi}{h^3}\int_{0}^{\infty}
\frac{p^2dp}{\exp{(E/kT)}+1}
\end{equation}
where p is the momentum of a particle, $E=\sqrt{p^2c^2+m_e^2c^4}$,
k is Boltzmann's constant, $h$ is Planck's constant and $m_e$, the
mass of an electron. For low temperatures such that $kT < m_ec^2$
this becomes:
\begin{equation}
n_\pm \approx 2(\frac{2 \pi m_ekT}{h^2})^{3/2} \exp{(-m_ec^2/kT)}
\end{equation}
\begin{displaymath}~~~~~~~~~~
=2.25\times 10^{30}(T_9/6)^{3/2} \exp{(-6/T_9)}~~cm^{-3}
\end{displaymath}
where $T_9=T/10^9$K.

We can express $dl$ in terms of changing redshift as
\begin{equation}
dl=|dr|(1+z) = 4R_g dz/(1+z)^2
\end{equation} for $R \approx 2R_g$
Substituting into Eq. 23 and using $T_\infty=T/(1+z)$ beyond the
photosphere we obtain the relation
\begin{equation}
1.77\times 10^{12}m
(T_{\infty,9}/6)\int_{\sqrt{T_{\infty,9}/6}}^{\sqrt{T_9/6}}e^{-6/T_9}d{\sqrt{T_9/6}}=
2/3
\end{equation}
Using Eqs. 16 and 19, we have numerically integrated this equation
to obtain the photosphere temperatures and redshifts for various
masses. The results are represented with errors below 1\% for
$1<m<10^{10}$ solar mass by the relations:
\begin{equation}
T_p=4.9\times 10^8 m^{-0.032}~~~~~~~~K
\end{equation}
and
\begin{equation}
1+z_p= 1840 m^{0.343}
\end{equation}
These relations, though little different from those of RL06,
correct an error in our previous development of the photosphere
temperature.

\medskip
\noindent{\bf D. Accretion Efficiency For MECO Surface
Luminosity}\\

We need to reconsider how accreting material interacts with a
MECO. BN06 assumed that for large redshifts, accretion energy
would be converted to luminosity immediately with 100\%
efficiency. In previous work (RL03, RL06) we mistakenly made the
same assumption. It is true that a MECO will eventually achieve
100\% efficiency, but the conversion takes place on the time scale
of the MECO radiative lifetime. Accreting particles that reach the
photosphere do not produce a hard radiative impact. They first
encounter soft photons, then photons of $\sim 1$ Mev and
electron-positron pairs near the photosphere and they eventually
penetrate the baryon surface where the net outflowing luminosity
is already at the local Eddington limit rate. This provides a very
soft landing.

Perhaps the easiest way to show the difference between landing on
a cushion of photons and striking a hard surface is to consider
the pressure that accreting matter could exert if stopped dead at
the photosphere and compare that to the radiation pressure already
there. Radiation pressure at the photosphere (hence the luminosity
also) would need to change by only about the same amount as the
accretion pressure for the MECO to remain stable.

Consider a particle of rest mass $m_o$ in radial free fall from
infinity. Its speed as it reaches the photosphere would be
essentially light speed, $c$, according to an observer at the
photosphere, and it would have a momentum of $\gamma m_oc$, where
$\gamma=1+z_p$ is the Lorentz factor\footnote{From the geodesic
equations of motion, the distant coordinate time, $t$, and proper
time, $\tau$, moving with the particle are related by
$dt=(1+z_p)^2 d\tau$. An interval $dt_p$ for an observer at rest
at the photosphere is related by $dt_p=\gamma d\tau=dt/(1+z_p)$,
from which it follows that $\gamma= 1+z_p$.}. If particles arrive
at the locally observed rate of $dN/dt_p$, then the quantity of
momentum deposited according to the observer at the photosphere
would be $(dN/dt_p)(1+z_p)m_oc$. With the substitution of
$dt/(1+z_p)$ for $dt_p$, this becomes $(dN/dt)m_o(1+z_p)^2c$. Thus
the rate of momentum transport, as observed at the photosphere is
$\dot{m_\infty}(1+z_p)^2c$, where $\dot{m_\infty}=m_odN/dt$.

If momentum were deposited uniformly at the photosphere, the
accretion pressure would be $p_{accr}= \dot{m_\infty}(1+z_p)^2
c/(4 \pi R^2)= 9.2\times 10^4 \dot{m_\infty} m^{-1.314}$ erg
cm$^{-3}$, using Eq. 27 for $z_p$ and $R=2R_g$. For the mass of
Sgr A* and $\dot{m_\infty}=1.3\times 10^{17}$ g/s, this would
yield $p_{accr}=2.9\times 10^{13}$ erg cm$^{-3}$. The radiation
pressure at the photosphere would be $p_{rad}=aT_p^4/3 = 1.5\times
10^{20} m^{-0.128}=2.2\times 10^{19}$ erg cm$^{-3}$, using Eq. 26.
This pressure ratio would be $p_{accr}/p_{rad}=1.3\times 10^{-6}$,
however, the actual ratio of accretion and radiation pressures
would be smaller yet because the entire momentum is not deposited
in a hard surface impact at the photosphere. Only about $10^{-7}$
of the incoming momentum is transferred to the entire pair
atmosphere. This is just too little to affect the radiation
leaving the photosphere. Finally, we note that in an equipartition
magnetic field, the temperature of the pair plasma layer is
buffered near the pair threshold temperature. Adding energy just
produces more pairs rather than raising the temperature. Of
course, for stability, the MECO must adjust its radiation rate to
accomodate a growing mass, but as distantly observed, the
luminosity produced by the incremental increase of mass is
negligible. For mass accretion rate $\dot{m_\infty}$ the radiated
luminosity increases at a rate of about
$(\dot{m_\infty}/M)L_\infty$. We conclude that the efficiency and
thermal equilibrium constraints of BN06 simply do not apply to the
MECO surface.

Eventually the radial infall of accreting particles ends, on
average somewhere below the MECO baryon surface. With the
temperature at the base of the pair atmosphere at $\sim 6\times
10^9$ K, Eq. 23 shows the local pair density to be near $10^{30}~
cm^{-3}$. The proper length thickness of the pair atmosphere is
$\sim 7\times 10^6~ cm$ and the optical depth is $\sim 5\times
10^{12}$. Assuming that protons in an accreting plasma interact
with the electron-positron pairs via coulomb scattering, the
fraction of the accretion energy that they can transfer directly
to the pair atmosphere is only $\sim 10^{-10}$. Photons in the
pair atmosphere are somewhat more effective in absorbing accretion
energy. For conditions in the pair atmosphere, about $10^{-7}$ of
the accretion energy can be removed by photon collisions with
incoming electrons, however, the escape cone allows only about one
in $10^{10}$ of the compton enhanced photons to escape. Most of
the brehmsstrahlung produced is therefore buried below the MECO
baryon surface which is itself covered by an extremely optically
thick layer of pairs for which the escape cone is negligibly
small. With due consideration of the small escape cone, it can be
easily shown that if accreting protons are stopped in a layer
below the photosphere for which the redshift is $z < z_s$, then
the ratio of accretion pressure to radiation pressure already in
the layer is $p_{accr}/p_{rad}=3.6\times 10^{-17}\dot{m}_\infty
/(m (1+z_s))=6.3\times 10^{-25}\dot{m}_\infty/m^{3/2}$, for
$1+z_s=5.67\times 10^7 m^{1/2}$. Accretion pressure is completely
insignificant even for accretion rates exceeding the classical
Edddington limit. In effect, a MECO just swallows accreting
particles about as effectively as a black hole.

\medskip
\noindent{\bf D.1 High/Soft Spectral States}\\
Reconsideration of the MECO surface accretion efficiency
necessitates a new interpretation of the high/soft spectral states
of disk accreting systems. At accretion rates above the transition
to the soft state, most of the luminosity of the high/soft state
must arise from the accretion disk rather than the MECO. This can
occur with high efficiency corresponding to an accretion disk that
can penetrate well inside what would otherwise correspond to the
marginally stable orbit of a non-magnetic black hole. The inner
disk of a MECO is supported in part by magnetic pressure, as
previously described in RL06 and disk accretion efficiencies can
reach 42\% at the photon sphere at $3R_g$. Inside the photon
sphere, gravitational redshift offsets the effect on luminosity of
additional radiant energy release by accreting matter.

\medskip
\noindent{\bf E. Electron Temperatures and Spectral Parameters}

The energy equation for the Bondi flow is
\begin{equation}
\frac{v^2}{2} + \frac{c_s^2}{\gamma_h-1} - \frac{GM}{r}=const=
\frac{c_{s,\infty}}{\gamma_h-1}
\end{equation}
where $v$ is the bondi flow speed, $\gamma_h$ the ratio of
specific heats for the gas, $M$ the mass of the central object and
$r$ the radial coordinate distance from its center. The continuity
equation for a rate of mass flow $\dot{m}$ is
\begin{equation}
\dot{m}=4\pi r^2 \rho v = const.
\end{equation}
We adopt the values $\dot{m}=3\times 10^{-6}~M_\odot /yr = 2\times
10^{20} ~g/s$, $n_\infty=\rho_\infty /m_p = 26~cm^{-3}$,
$c_{s,\infty} = 5.5 \times 10^7 ~ cm/s$ (Baganoff 2003) and a
specific heat ratio of $\gamma_h=5/3$ for atomic hydrogen. Having
adopted these values, they set the Bondi radius as $1.63\times
10^{17}~cm$ and they must be entered consistently in the energy
equation. We use $c_s=\gamma_h K \rho^{(\gamma_h-1)}$, where
$K=p_\infty /\rho_\infty^{\gamma_h}$ for an adiabatic flow. Using
$\rho=\dot{m}/(4\pi r^2 v)$ and substituting into the energy
equation produces an equation of the form
\begin{equation}
v^2+Av^{-2/3}/r^{4/3}-B-C/r = 0
\end{equation}
where A, B and C are constants. Solutions for v for various values
of r are given in Table 2. Once v has been determined, $\rho$ can
be obtained from the continuity equation, and the pressure and
proton temperature can be found from the adiabatic conditions. For
$\gamma_h=5/3$, the inflow never becomes supersonic, but it very
closely approaches sound speed (which is half what the free-fall
speed would be) by the time the magnetosphere is reached.

For the dilute plasma at the Bondi radius, there is a relatively
low frequency of electron-proton collisions. The kinetic energy
gained by falling electrons is not sufficient to keep them in
thermal equilibrium with the ions. Approximately half of the
gravitational energy released goes into heating the ions while the
electrons remain relatively cold. The electrons do gain some
energy, which we can calculate, in collisions with protons. As the
plasma falls by $dr$, the time taken is $dr/v$. An electron
travels $v_edt = v_edr/v$ in this time. The number of collisions
with ions in this time is the distance travelled divided by the
mean free path, $1/n\sigma_c$, where $\sigma_c= (\pi
e^4/E_e^2)\ln{\Lambda}$ is the coulomb cross-section.
$E_e=(3/2)kT_e$ is the electron energy, $T_e$ the electron
temperature and $\ln{\Lambda}$ is the coulomb logarithm, which we
assume to have the value 30. For collisions with protons which, on
average, have much more momentum than the electrons, the fraction
of proton energy, $(1/2)m_pv^2$, that can be transferred is
approximately $2v/v_e$. Then while falling dr, an electron gains
energy $dE_e=(3/2)kdT_e =n \sigma_c m_p v^2dr$. Taking
$v=(GM/2r)^{1/2}$ for both the proton thermal speed and speed of
descent and substituting for  $\sigma_c ~(\propto T_e^{-2})$ and
$n~ (\propto r^{-3/2})$, this integrates to yield $T_e \propto
r^{-1/2}$ and specifically,
\begin{equation}
T_e=9.3\times 10^{15} r^{-1/2}~~~ K
\end{equation}
yields the values in Table 2.

An electron moving with the protons at speed $v~ (=dr/dt)$ in a
magnetic field would emit cyclotron radiation at a rate
$dE_R/dt=2e^4v^2B^2/(3c^5m_e^2)$. Comparing this rate for the
axial dipole field of a MECO with the mass of Sgr A* with
$dE_e/dt$ above, we find that the radiation rate would exceed the
rate that electron energy is extracted from the protons for $B >
8600 ~G$, which holds for $r<5\times 10^{13}~cm$, and cyclotron
frequencies above $14~ GHz$. Nearer the MECO at smaller axial
distances and higher frequencies, the electron energy extracted
from the protons in the flow is essentially all that is available
for radiation. This permits a rough calculation of the luminosity
generated in the polar inflow. We approximate the polar inflow
geometry within the magnetosphere as a cone of lateral radius
$a=r_cz/z_m(in)$ at distance z from the magnetic pole. See Figure
1. The number of electrons in the cone in axial thickness $dz$ is
$dN=n \pi a^2 dz$. Thus the rate that they can radiate is about
$dL=dN dE_e/dt=n\pi (r_c/z_m(in))^2z^2n \sigma_c m_pv^3dz$.
Substituting for $n(z)$, $\sigma_c(z)$, $v(z), T_e(z)$ as before
and using $r_c=3.65\times 10^{13}~cm$, $z_m=10^{15}~cm$ from
equations in Table 1, this reduces to $dL=1.1\times 10^{41}
z^{-3/2} dz$, which integrates to give the luminosity for polar
inflow to axial distance $z$ as
\begin{equation}
L=4.2\times 10^{41} z^{-1/2}~~~ erg/s
\end{equation}
This last result includes an additional factor of two to account
for inflow into both magnetic poles.

Where the flow is optically thin above $10^{12}~Hz$, the Larmor
frequency should strongly dominate the spectrum. This frequency is
a function of the magnetic field strength, which varies as
$r^{-3}$. Alternatively, we can say that position $z$ in the polar
axial inflow is correlated with the dominant radiation frequency
as $z \propto \nu^{-1/3}$. Substituting $1.36\times 10^{17}
\nu^{-1/3}$ for $z$ in the expression just preceding Eq. 32, one
obtains
\begin{equation}
dL =2\times 10^{32} \nu^{-5/6} d\nu~~~~~ erg/s
\end{equation}
hence the spectral index in the optically thin sub-mm/NIR would be
$-0.83$ for the MECO model.

\end{document}